\documentclass[manuscript]{aastex}
\usepackage{lineno}
\usepackage{tikz}
\usepackage{amsmath}
\usepackage{hyperref}

\usepackage{graphicx}

\usepackage{pdflscape}
\usepackage[normalem]{ulem} 
\usepackage{soul} 
\usepackage{aas_macros}

\slugcomment{\textbf{2026 February 4} }

\shorttitle{41P}
\shortauthors{Jewitt}

\begin{document}

\title{Reversal of Spin: Comet 41P/Tuttle–Giacobini–Kresak}

\author{
David Jewitt$^{1}$ 
} 
\affil{$^1$Department of Earth, Planetary and Space Sciences, UCLA, 595 Charles Young Drive, Los Angeles, CA 90095\\}

\email{djewitt@gmail.com}

\begin{abstract}
The rotations of cometary nuclei are known to change in response to outgassing torques. The nucleus of Jupiter family comet 41P/Tuttle–Giacobini–Kresak exhibited particularly dramatic rotational changes when near perihelion in 2017 April.  Here, we use archival Hubble Space Telescope observations from 2017 December to study the post-perihelion lightcurve of the nucleus and to assess the nucleus size.  From both Hubble photometry and non-gravitational acceleration measurements we find a diminutive nucleus with effective radius $r_n$ = 500$\pm$100 m.  Systematic optical variations are consistent with a two-peaked (i.e., rotationally symmetric) lightcurve with period  0.60$\pm$0.01 days, substantially different from periods measured earlier in 2017.  The spin of the nucleus likely reversed between perihelion in 2017 April and  December as a result of the outgassing torque. We infer a dimensionless moment arm $k_T$ = 0.013, about twice the median value in short-period comets.  The lightcurve range of 0.4 magnitudes indicates a projected nucleus axis ratio $\gtrsim$1.4:1, while the active fraction of the nucleus decreased from  $\sim$2.4 in 2001 (suggesting augmentation of the gas production by sublimating coma ice grains) to $\sim$0.14 in 2017, a result of long-term modification of the surface.  We find that the physical lifetime of this small nucleus to spin-up is short compared to the reported $\sim$1500 year dynamical time spent in the current orbit.  Two limiting reconciliations of this inequality  are suggested.  The nucleus could be in a state of unusually strong  activity, leading us to over-estimate the average mass loss rate and outgassing torque and so to under-estimate the physical lifetime.  Alternatively, the nucleus could be the surviving remnant of a once larger body for which outgassing torques were less effective in changing the spin.

\end{abstract}


\section{INTRODUCTION}
\label{intro}
Jupiter family comet 41P/Tuttle–Giacobini–Kresak (hereafter TGK) has orbital semimajor axis $a$ = 3.085 au, eccentricity $e$ = 0.661 and inclination $i$ = 9.2\degr.  Perihelion and aphelion lie at 1.045 au and 5.124 au, respectively.  Despite repeatedly dipping into Jupiter's Hill sphere when near aphelion, numerical integrations of the motion suggest that TGK's orbit is relatively stable.  The comet was injected to its present orbit following the last close encounter with Jupiter $\sim$1,500 years ago and it will maintain an orbit similar to the present one for perhaps the next $10^4$ years \citep{Poz18}. The likely source region, as for Jupiter family comets in general, is the Kuiper belt and the long-term fate, if the nucleus survives long enough, is to be scattered into the Sun or ejected from the solar system.

TGK was particularly well placed for observation in its 2017 apparition (perihelion date was 2017 Apr 12.7, or Day of Year in 2017, $t_{DOY}$ = 103) when the water production rate was a modest $Q_{OH} \sim 3\times10^{27}$ s$^{-1}$ ($\sim$90 kg s$^{-1}$; \cite{Com20}.  The rate was an order of magnitude larger two orbits before, in 2001  \citep{Mou18},  indicating evolution of the active fraction of the nucleus from orbit to orbit. TGK has also shown photometric outbursts both large (sudden brightening by 9 magnitudes in 1973; \cite{Kre74}) and small (0.6 magnitudes in 2017; \cite{Boe20}), reflecting instability of the nucleus surface.  Remarkable observations from 2017 show a rapid increase of the rotation period of the nucleus, deduced both photometrically \citep{Bod18} and from rotating jets  (\cite{Far17}, \cite{Sch19}). The period more than doubled, from $P \sim$20 hours to $\sim$53 hours, over the course of two months near perihelion. The simplest explanation of the changing period is that the nucleus was torqued by recoil forces from anisotropic outgassing, as has been widely demonstrated in other comets \citep{Jew21}. Indeed, ground based observations set an upper limit to the nucleus radius $r_n \lesssim$ 0.7 km (visual geometric albedo 0.04 assumed; \cite{Tan00}), a size which renders the nucleus susceptible to rapid spin evolution through outgassing torques. An explanation incorporating excited (non-principal axis, or NPA) rotation might exist \citep{How18}, however, no NPA model has been presented  and NPA rotation may be difficult to reconcile with the simple morphology of the coma jets \citep{Sch19}. In this work, as in \cite{Bod18} and \cite{Sch19}, we  adopt Occam's Razor as our starting point, and assume that the nucleus is in principal axis rotation.  

In this paper, we analyze unpublished observations of TGK from the Hubble Space Telescope (HST), revealing continued, post-perihelion evolution of the spin period and placing independent constraints on the size of the nucleus.

\section{OBSERVATIONS}

Our data consist of archived observations  from the HST General Observer program GO 15421 (PI: T. Farnham), taken about eight months after perihelion in the period UT 2017 December 11 - 14 ($t_{DOY}$ = 345 - 348).  As shown in Table \ref{geometry}, the comet was then outbound at heliocentric distance $r_H \sim$ 2.79 au. The WFC3 camera, with pixels 0.04\arcsec~on a side, was used together with the ultra-wide F350LP filter, which has a cut-on wavelength near 3500\AA, a central wavelength  near 5874\AA~and a full width at half maximum (FWHM) of $\sim$4800\AA. (Other filters having smaller FWHM were used in GO 15421 but gave lower signal-to-noise ratios leading us to exclude them from the present study). Each WFC3 pixel projects to $\sim$96 km at the $\Delta \sim$ 3.3 au distance to the comet, giving a  Nyquist sampled (two pixel) resolution of 192 km.  The WFC3 images were read out as subframes of 1030$\times$1087 pixels, corresponding to a field of view about 40\arcsec~square. The telescope was tracked to follow the non-sidereal motion of TGK, causing background stars and galaxies to trail during the 160 s integrations.  Images of the comet that were found to be affected by trailed field objects and/or by cosmic ray strikes were rejected from consideration, leaving 24 useful images taken on the four consecutive days.

\section{DISCUSSION}

\begin{figure}
\epsscale{0.99}

\plotone{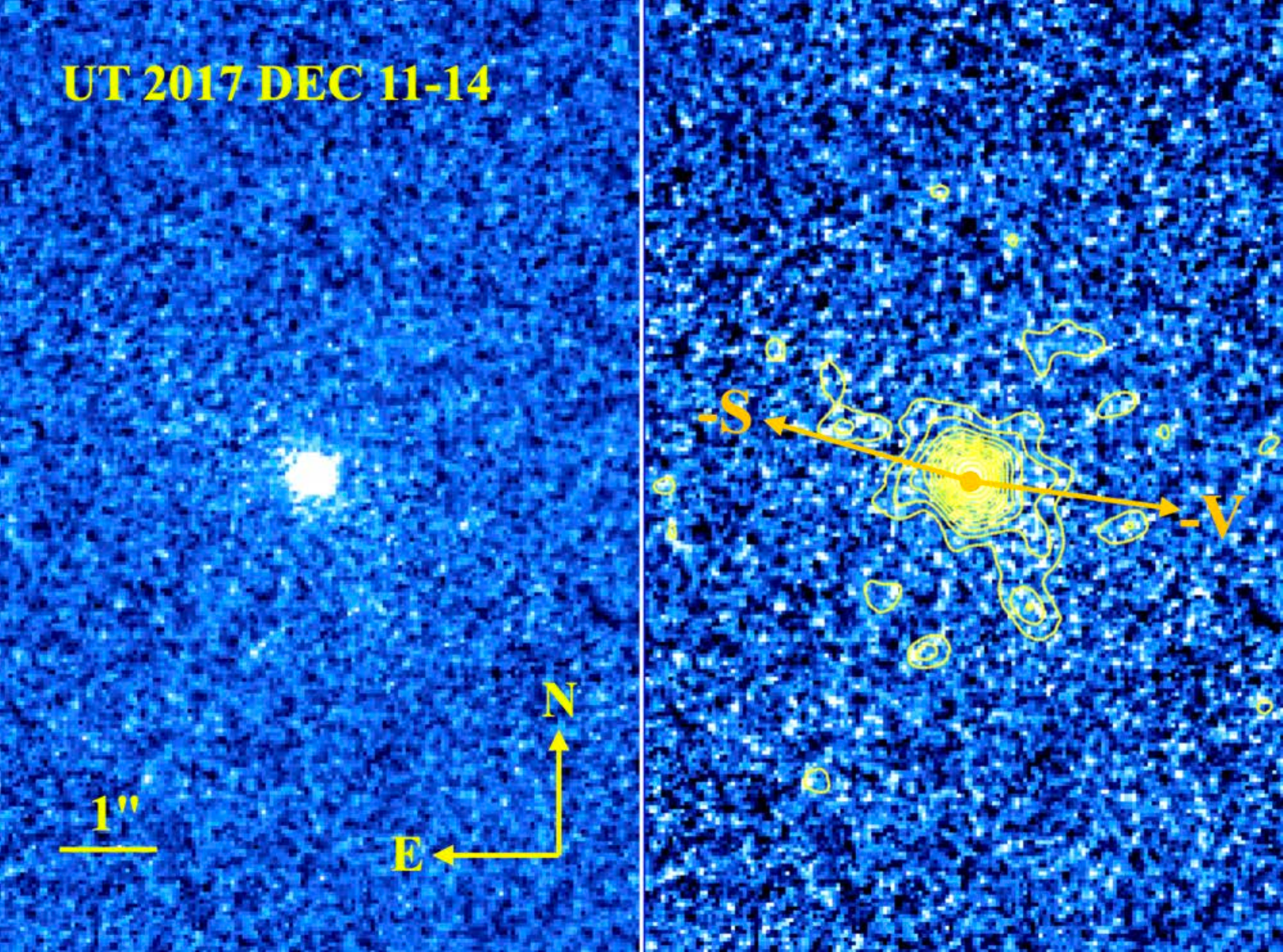}

\caption{(Left panel:) Composite 3840 s integration on TGK.  A 1\arcsec~scale bar and the marked cardinal directions apply to both panels.  (Right panel:) Same image contoured to emphasize the near-nucleus coma.  Direction arrow show the antisolar direction (-S) and the projected negative heliocentric velocity vector (-V).  \label{image}}
\end{figure}

We formed  a stacked image  from the 24 exposures, with a combined integration time of 3840 s (Figure \ref{image}). The stacked image is nearly point-like, but shows barely perceptible diffuse emission around TGK several arcseconds in extent and possibly asymmetric, with an axis near position angle 250\degr~to 260\degr.  This direction roughly matches the projected negative heliocentric velocity vector direction (262\degr) and is opposite to the anti-solar direction (marked with arrows in the figure and also see Table \ref{geometry}).  Large, slow-moving particles are expected to concentrate near the projected orbit, but the signal-to-noise ratio in the coma is so low that it is impossible to draw any strong conclusion about the nature of the coma particles from the measured position angle.  Likewise, we cannot usefully measure the radial surface brightness profile of the coma because it is so faint.  Even if the coma had been brighter the small angle ($\sim$2\degr; Table \ref{geometry}) between HST and the orbital plane of TGK causes  syndynes and synchrones to overlap in projection, so limiting the utility of dust dynamics models.

\subsection{Size of the Nucleus}

We obtained photometry using circular apertures of angular radius 5 and 10 pixels (0.2\arcsec~and 0.4\arcsec) respectively, with sky subtraction from the median count within a concentric annulus having inner and outer radii 10 pixels and 110 pixels (0.4\arcsec~and 4.4\arcsec).  From the photometric scatter we estimate a single image  uncertainty in the 5 pixel aperture of $\sigma_V$ = 0.07 magnitudes, slightly larger than the 0.05 magnitudes predicted by the WFC3 Exposure Time Calculator\footnote{\url{https://etc.stsci.edu/etc/input/wfc3uvis/imaging/}}.  The larger value of the empirical uncertainty is attributed to the spatial structure and image-to-image variability of the sky background, neither of which can be accounted for in the prediction software.

The  photometry provides an independent estimate of the effective nucleus radius, $r_n$, defined as the radius of a circle having the same scattering cross-section as the nucleus.  The mean and median five pixel magnitudes from Table \ref{photometry} are $\overline{V_5} = 24.18\pm 0.03$ (where the uncertainty reflects mainly the rotational variations of TGK) and  24.15, respectively. Taking the latter value and correcting to $r_H$ = $\Delta$ = 1 au using the inverse square law, and to $\alpha$ = 0\degr~assuming a phase coefficient 0.04 magnitudes degree$^{-1}$, we find absolute magnitude $H_5$ = 18.69$\pm$0.03. The quoted uncertainty ignores systematic error due to the unmeasured phase function (which, in turn, may adopt different values for the nucleus and for the dust in the coma).  At phase angle 16\degr, a $\pm$0.01 magnitude degree$^{-1}$ error in the phase coefficient would produce a $\pm$0.16 magnitude error in $H_5$. However, $H_5$  still  over-estimates the nucleus brightness because of the effects of near nucleus coma.  To estimate the magnitude of the coma contamination, we compare photometry within the 5 and 10 pixel radius apertures.  For TGK, the mean magnitude difference between these apertures is $\Delta V$(5-10) = 0.28$\pm$0.02 while for point sources observed with WFC3 and the 350LP filter, we measured  $\Delta V_{\star}$(5-10) = 0.081$\pm$0.003.  The difference, $\Delta V$(5-10) $-$ $\Delta V_{\star}$(5-10)= 0.20$\pm$0.02 magnitudes, is a measure of the coma in the 5 to 10 pixel annulus.  In a canonical steady state coma, the encircled flux is proportional to the aperture radius.  If so, the  magnitude of the coma contained in the 5 to 10 pixel annulus (about 0.2 magnitudes) is equal to the magnitude of the coma in the central 5 pixel aperture. In this case, our best estimate of the nucleus absolute magnitude is $H_5$ = 18.89$\pm$0.03.    

The cross-section, $C_e$, and the absolute magnitude are related by the inverse square law

\begin{equation}
p_V C_e = 2.25\times10^{22} \pi 10^{-0.4(H - V_\odot)},
\label{invsq}
\end{equation}

\noindent where $V_{\odot}$ = -26.74 is the apparent V magnitude of the Sun at 1 au \citep{All73}.  Substitution, with $p_V$ = 0.04, gives effective nucleus cross section $C_e = (9.9\pm0.3)\times10^5$ m$^2$, and the radius of an equal area nucleus is $r_n = (C_e/\pi)^{1/2}$ = 560$\pm$10 m, where the quoted uncertainty propagates only the statistical error on $H_5$.   This value is smaller than but consistent with the published estimate based on ground-based photometry, $r_n \le$ 700 m  \citep{Tan00}, for which larger coma contamination might be expected as a result of the poorer ground-based resolution.   

Radar observations at 12.6 cm wavelength  and at unknown rotational phase  set a lower limit to the nucleus radius $r_n \gtrsim$ 450 m  \citep{How17}, which is compatible with the value estimated optically from Equation \ref{invsq}. However, we note that the radar size is dependent on the assumption of an unmeasured radar albedo and that the radar observations have unfortunately not been published other than as an abstract.

A third and independent estimate of the nucleus size may be obtained from the non-gravitational acceleration of TGK.  The JPL Horizons Small Body Lookup\footnote{\url{https://ssd.jpl.nasa.gov/tools/sbdb_lookup.html\#/}} gives the non-gravitational acceleration parameters 

\begin{equation}
A1 	= 6.6\times10^{-9} \pm 	5.8\times10^{-11} \textrm{ au day}^{-2} \textrm{ and }
A2 	= -2.1\times10^{-10} \pm 	7.7\times10^{-11} \textrm{ au day}^{-2}.
\end{equation}

\noindent Combining these components in quadrature, and noting that 1 au day$^{-2}$ is equal to 20.1 m s$^{-2}$, the total acceleration on the nucleus is $\zeta = 1.3\times10^{-7}$ m s$^{-2}$, almost entirely in the direction radial to the Sun (given by component A1).  Interpreted as recoil acceleration, we calculate the nucleus radius from  \cite{Jew20}

\begin{equation}
r_n = \left(\frac{3 k_R \dot{M} V_{th}}{4 \pi \rho_n \zeta}\right)^{1/3}
\label{recoil}
\end{equation}

\noindent in which $k_R$  reflects the anisotropy of the sublimation ($k_R$ = 0 for isotropic ejection, $k_R$ = 1 for perfectly collimated flow along the comet-Sun line).  We substitute $k_R$ = 0.5 (calculated from the measured acceleration and mass loss parameters of 67P/Churyumov-Gerasimenko, see Appendix to \cite{Jew20}), $\dot{M}$ = 90 kg s$^{-1}$ at perihelion (\cite{Com20}, \cite{Mou18}), $V_{th}$ = 500 m s$^{-1}$ as the thermal speed in the coma near 1 au, $\rho_n$ = 500 kg m$^{-3}$ \citep{Gro19}, to find $r_n \sim$ 440 m.  The recoil model has limitations (e.g., some of the gas production might come from icy grains in the coma, so reducing the mass loss rate from the nucleus and leading to a smaller radius by Equation \ref{recoil}).  However, the agreement between radius estimates from photometry (Equation \ref{invsq}) and from Equation \ref{recoil} is encouraging, especially considering that neither $p_V$ nor $k_R$ nor $\rho_n$ have been measured in TGK.  

We conclude that the available data from optical, radar and non-gravitational acceleration measurements confirm that the nucleus of TGK is a sub-kilometer object, with an effective circular radius 440 m to 560 m.  As a working value, we take $r_n$ = 500$\pm$100 m in the remainder of this paper.

\subsection{Nucleus Rotation}

\begin{figure}
\epsscale{0.99}

\plotone{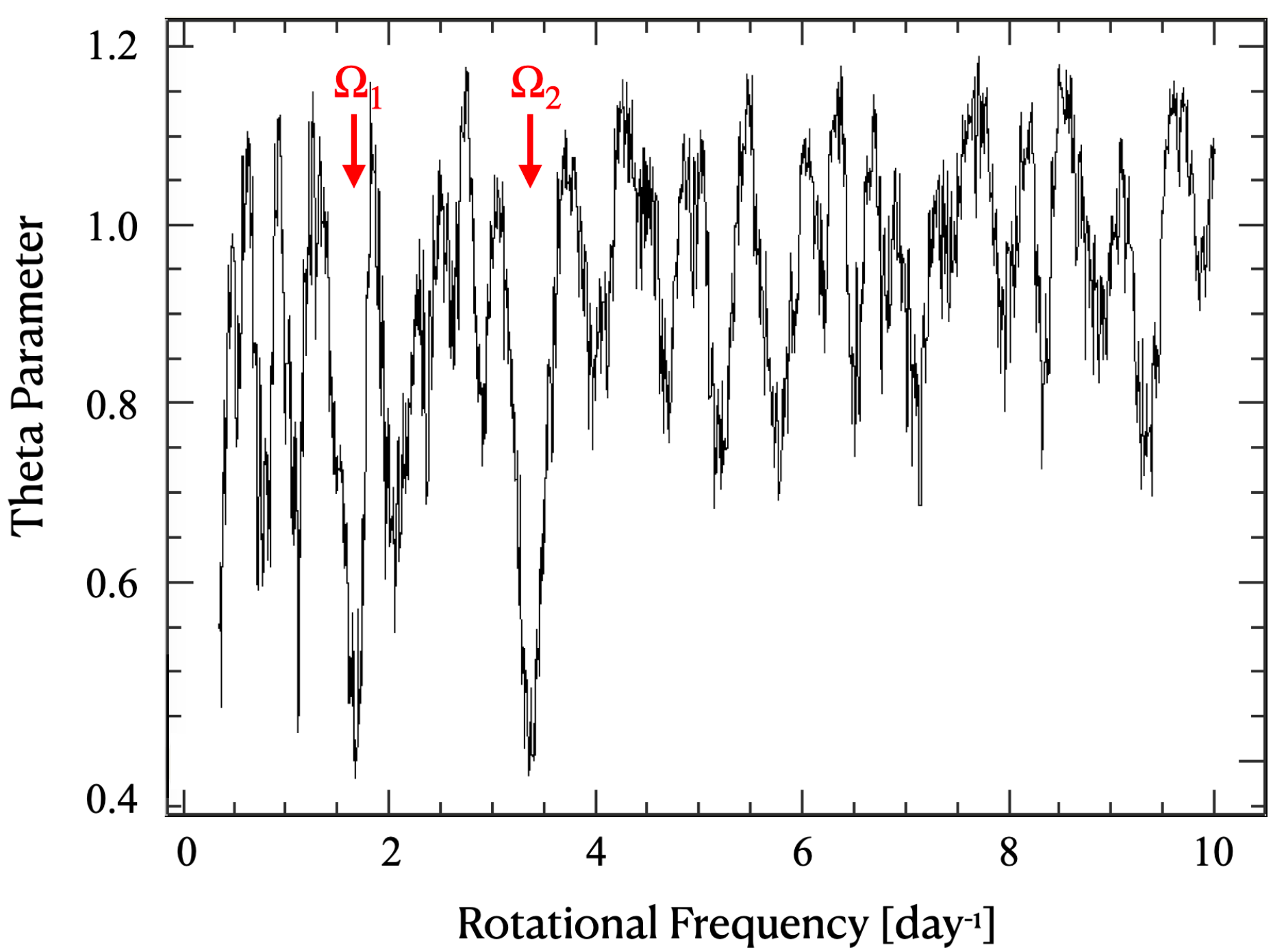}

\caption{Phase dispersion minimization plot showing the minimization parameter as a function of the assumed rotational frequency.  Minima at $\Omega_1$ = 1.669 day$^{-1}$ and $\Omega_2$ = 3.339 day$^{-1}$ correspond to rotational periods at  0.599 day and  0.299 day, respectively.   
$\Omega_1$ and $\Omega_2$ give two-peaked and one peaked lightcurves.  \label{PDM}}
\end{figure}

\begin{figure}
\epsscale{0.99}

\plotone{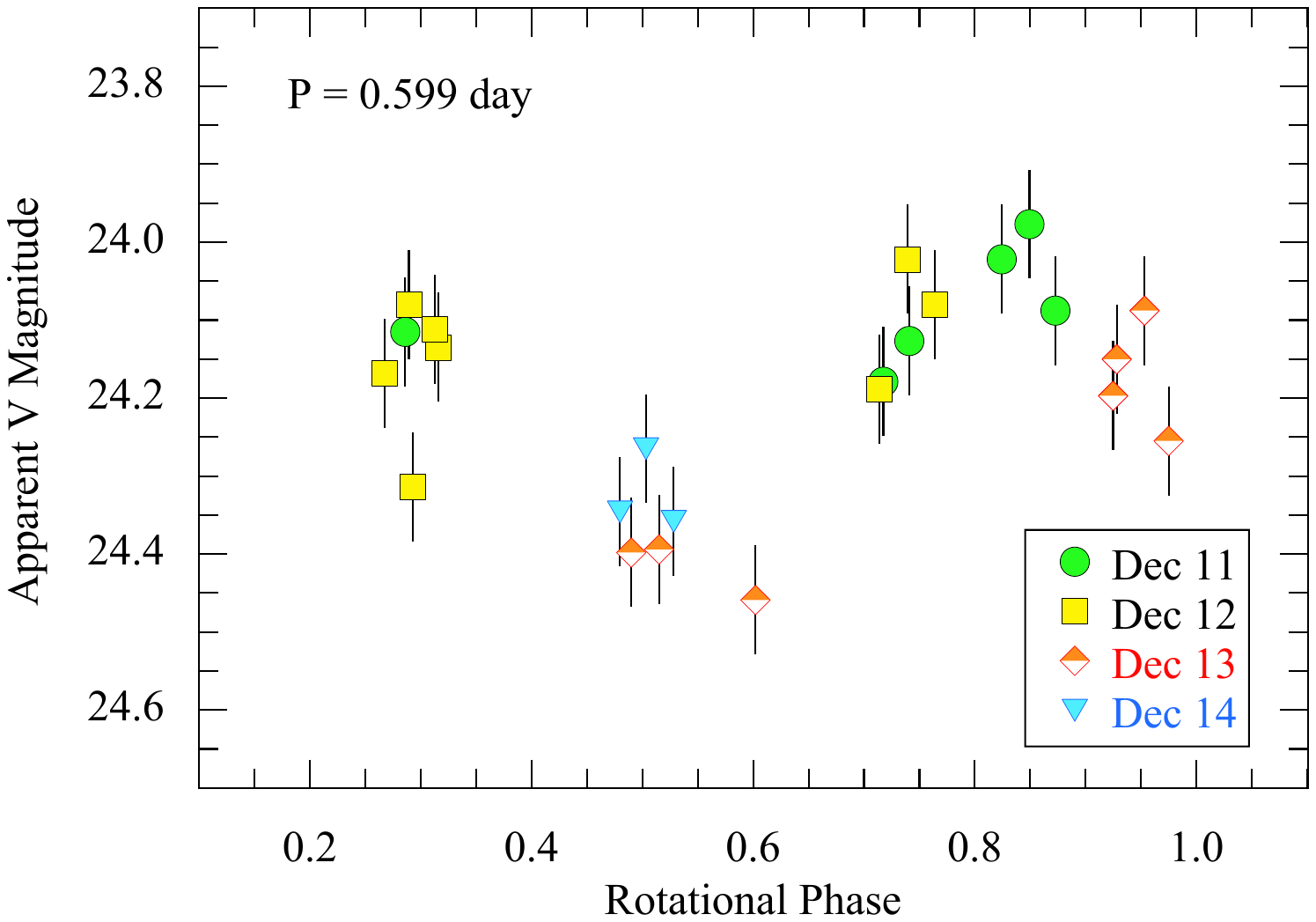}

\caption{Two-peaked lightcurve computed for the period $P$ = 0.599 day (14.4 hour), corresponding to $\Omega_1$ in Figure \ref{PDM}.  Data are distinguished by the UT 2017 December date on which they were obtained, as marked. \label{lightcurve}}
\end{figure}

The photometry in Table \ref{photometry} shows a variation of about $\Delta V$ = 0.4 magnitudes, significantly larger than the $\pm$0.07 magnitudes uncertainty on individual measurements.  We used phase dispersion minimization (PDM; \cite{Ste78}) to search for periodicity in the data. Figure \ref{PDM} shows a plot of the PDM quality metric, $\Theta$, against the inverse rotation period, $\Omega = 1/P$, where $P$ is the rotation period in days.  The data are sparse and non-uniformly sampled, giving rise to aliases in the PDM plot.  However, while there are several minima in the figure, only the two deepest, at rotational frequencies $\Omega_1$ = 1.669 day$^{-1}$ and $\Omega_2$ = 3.338 day$^{-1}$, give plausible lightcurves.  (The uncertainties on $\Omega$ were estimated at $\pm$0.015 day$^{-1}$ by comparing lightcurves computed for different inverse periods). The smaller frequency, $\Omega_1$,  corresponds to a two-peaked lightcurve with period $P_1$ = 0.599 day) while the larger frequency, $\Omega_2 = 2\Omega_1$, gives a single peaked lightcurve with $P_2$ = 0.300 day.  Most lightcurves of small solar system bodies are two-peaked because of rotational symmetry. Therefore, we identify $P_1$ in Figure \ref{PDM}  as the best estimate of the nucleus rotation period (see phased lightcurve in Figure \ref{lightcurve}).   It is noteworthy that we do not find acceptable lightcurve solutions in the 24 to 48 hour period  (0.5 $\le \Omega \le 1$ day$^{-1}$) range that is prominent in the near-perihelion observations by \cite{Bod18} and \cite{Sch19}.  The $\Delta V$ = 0.4 magnitude range of the lightcurve (Figure \ref{lightcurve}) corresponds to a nucleus whose axis ratio projected into the plane of the sky is $a/b \gtrsim$ 1.4, a typical axis ratio for comets and small asteroids.

\subsection{Secular Variations}
The torque responsible for the slowing rotation observed in early 2017 (\cite{Bod18} and \cite{Sch19}), if sustained, should eventually stop the rotation ($\Omega \rightarrow 0$) and thereafter spin the nucleus in the opposite direction.  Lightcurve data by themselves cannot distinguish between prograde and retrograde rotation. Therefore, we cannot know if the rotation in 2017 December has the same sense as in the near-perihelion observations, or whether the rotation has reversed in the intervening period.  In other words, inverse periods $\Omega_1$ = +1.669 day$^{-1}$ and $\Omega_1$ = -1.669 day$^{-1}$ are both consistent with the photometric data.  However, we argue that reversal is the more likely of these two possibilities, as we now describe.  

\begin{figure}
\epsscale{0.8}

\plotone{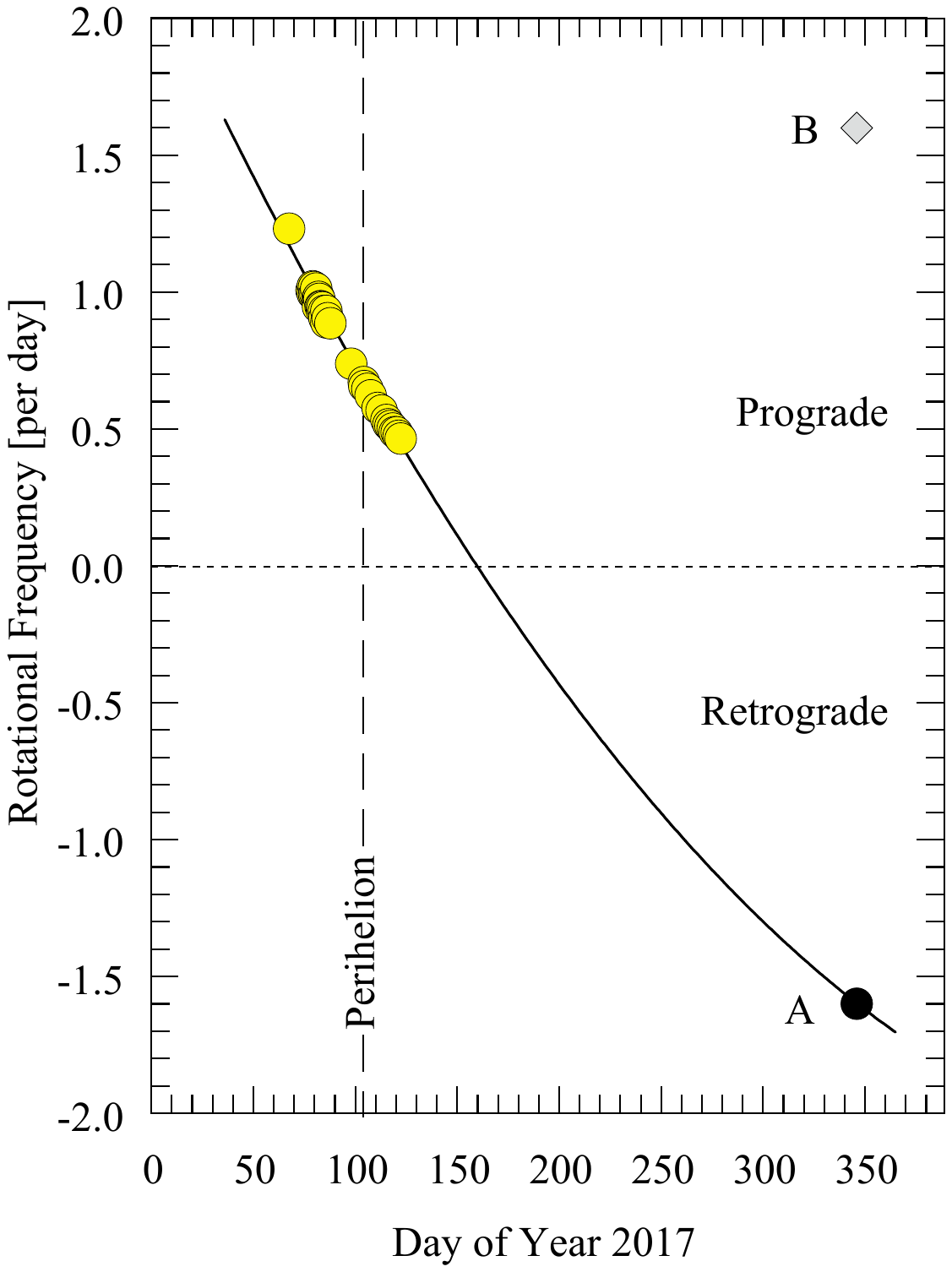}

\caption{Rotational frequency as a function of time expressed as Day of Year in 2017. The yellow-filled circles show data from \cite{Bod18} and \cite{Sch19}.  Points A (black-filled circle) and B (grey diamond) show the two solutions for prograde and retrograde rotation deduced from the HST lightcurve.  The solid black line is a parabola added to guide the eye.  Frequencies above(below) the dashed horizontal line are prograde(retrograde).  The date of perihelion is indicated by a dashed vertical line. \label{freq}}
\end{figure}

The available rotation measurements are compared in Figure \ref{freq}, where have plotted the new HST measurement twice, with Point A corresponding to a reversal of spin and Point B corresponding to rotation in the same sense as near perihelion in 2017. Point A fits on a smooth extrapolation from the near-perihelion data.  We show an (arbitrary) second order polynomial fit to emphasize the smooth continuation of the decreasing $\Omega$ first reported by \cite{Bod18} and \cite{Sch19}. The fit requires that $d\Omega/dt$ should decrease as $r_H$ increases, which is to be expected if sublimation is the source of the torque.  The fit gives $\Omega$ = 0 day$^{-1}$ on $t_{DOY} =$  160 (UT 2017 June 9) and the gradient of the plotted curve reaches $d\Omega/dt$ = 0 on $t_{DOY}$ = 542, corresponding to 2018 June.  At this time, still 18 months from aphelion (UT 2019 December 29), the predicted rotational frequency would have been $\Omega$ = -2.17 day$^{-1}$ ($P$ = - 0.46 day).  If instead we assume that Point B in Figure \ref{freq} is the correct solution for $\Omega$, then the outgassing torque must have suddenly reversed for unknown reasons after the \cite{Bod18} and \cite{Sch19} observations.  While we cannot prove that this did not happen, it seems more natural to assume a smooth continuation of the near-perihelion deceleration passing through Point A.  The possibility of reversal of the spin was discussed by \cite{Bod18} (see their Figure 4) but it appears to have occurred one orbit before their prediction time.

The spin deceleration is likely to be due to torques arising from asymmetric outgassing. Following the order of magnitude derivation in \cite{Jew21}, we consider the result of an outgassing torque applied to a spherical nucleus of radius $r_n$, density $\rho_n$ and rotating with period $P$ (frequency $\Omega = 1/P$).  The angular momentum of such a nucleus is $L = 2\pi k_I  M_n r_n^2  \Omega$, where $M_n = 4\pi r_n^3 \rho_n/3$ is the nucleus mass and the coefficient of inertia for a uniform  density sphere is $k_I$ = 2/5.  Combining, we obtain $L = 16\pi^2 \rho_n r_n^5 \Omega/15$.  The nucleus is acted upon by a torque of magnitude $T = k_T  V_{th} r_n \dot{M}$, where $k_T$ is the dimensionless moment arm, a constant equal to the fraction of the outflow momentum exerted non-radially, $\dot{M}$ is the rate of loss of mass from the nucleus and $V_{th}$ is the average speed of the material leaving the nucleus.  Recognizing that $T = dL/dt$, we write

\begin{equation}
\dot{\Omega}= \pm \left(\frac{15 k_T V_{th}}{16\pi^2 \rho_n r_n^4}\right) \dot{M}
\label{dodt}
\end{equation}

\noindent for the rate of change of the angular velocity of the nucleus in response to outgassing at rate $\dot{M}$.  The sign on the right hand side of Equation \ref{dodt} expresses  whether the torque acts to accelerate or decelerate the rotation. The evidence (Figure \ref{freq}) shows that $\dot{\Omega} < 0$ in 2017. The evolution of the angular velocity is then obtained by integrating Equation \ref{dodt} with respect to time as

\begin{equation}
\Omega(t) = \Omega_0 \pm \left(\frac{15 k_T V_{th}}{16\pi^2 \rho_n r_n^4}\right)\int_{t_0}^t \dot{M}(r_H(t)) dt.
\label{Omega}
\end{equation}

\noindent Here, $\Omega_0$ is the angular frequency at initial time $t_0$ and the mass loss rate is recognized as a function of heliocentric distance, $r_H$, which is itself a function of elapsed time, $t$.

Ideally, we could use measurements of  $\dot{M}(r_H(t))$ to compute the integral in Equation \ref{Omega} and infer the evolution of $\Omega(t)$.  However, in the case of TGK we possess no useful data on $\dot{M}(r_H(t))$ outside of a three week window near perihelion in Lyman $\alpha$ measurements by \cite{Com20}.  Instead, we use Equation \ref{dodt} and local measurements of $\Omega(t)$ and $\dot{M}(r_H(t))$ to estimate the moment arm, $k_T$, for the nucleus of TGK, from

\begin{equation}
k_T = \frac{16\pi^2 \rho_n r_n^4 \Omega(t)}{15  V_{th} \dot{M} }.
\label{k_T}
\end{equation}

\noindent For example, the polynomial  fit shown in Figure \ref{freq} gives  $\dot\Omega$ = -0.0131 day$^{-2}$ (equal to -1.75$\times10^{-12}$ s$^{-2}$) at perihelion.  Again using $\dot{M}(r_H(t))$ = 90 kg s$^{-1}$ at this time, and substituting for $V_{th}$, $\rho_n$ and $r_n$ from above, Equation \ref{k_T} gives $k_T$ = 0.013. The uncertainty on $k_T$ is both considerable and difficult to estimate.  For example, $k_T$ is particularly sensitive to the nucleus radius, $r_n$. With $k_T \propto r_n^4$ (c.f., Equation \ref{k_T}), a $\pm$20\% uncertainty in $r_n$ translates to $\pm$80\% uncertainty on $k_T$.  (This $r_n^4$ sensitivity largely explains why \cite{Jew21}, who used $r_n$ = 700 m from \cite{Tan00}, deduced $k_T$ = 0.036 for TGK, about three times larger than found here.)
The new estimate is the same as the moment arm for comet 2P/Encke, and about twice the median value ($k_T$ = 0.007) determined for a sample of eight short-period comet nuclei \citep{Jew21}, but still lies within the range of $k_T$ values measured in that work. Rotational changes in TGK, although dramatic by the standards of most other cometary nuclei, are a simple consequence of its small size, not of outgassing that is unusual in magnitude or angular pattern.

\subsection{Active Fraction} 

In thermal equilibrium with sunlight, a perfectly absorbing spherical water ice nucleus at 1 au would sublimate from the sunward hemisphere at the average rate $f_s \sim 2\times10^{-4}$ kg m$^{-2}$ s$^{-1}$.  To sustain the water production rate measured at perihelion in 2017, namely $\dot{M}$ = 90 kg s$^{-1}$ (\cite{Com20}, \cite{Mou18}), would require a sublimating ice area $A_{H_2O} = \dot{M}/f_s$ which, by substitution, gives $A_{H_2O} = 4.5\times10^5$ m$^2$ (0.45 km$^2$).  This compares with the surface area of a 0.5$\pm$0.1 km nucleus, $A_n = 4\pi r_n^2$ = 3.1$_{-1.1}^{+1.4}$ km$^2$, and gives an active fraction $f_A = A_{H_2O}/A_n \sim$ 0.14$_{-0.04}^{+0.09}$.   The sustained water production rate from TGK was $\sim$1500 kg s$^{-1}$ in 2001 \citep{Com20}, giving $f_A \sim$ 2.4$_{-0.8}^{+1.4}$ then and leading  the nucleus of TGK to be categorized as ``hyper-active''\footnote{It could be argued that, since sublimation is almost exclusively confined to the hot Sun-facing hemisphere of the nucleus,  any value  $f_A >$ 1/2 should qualify a nucleus as being hyper-active, but the conventional definition is $f_A >$ 1.}.  Hyper-activity, when present, can result from the sublimation of icy grains in the coma.  Changes in $f_A$ and reported observations of large photometric outbursts in earlier orbits \citep{Kre74} are indicators of surface instability.  It is natural to expect that other nucleus parameters, notably $k_R$ and $k_T$, may likewise be strong functions of time.

\subsection{Lifetime}

TGK entered its current orbit $\sim$1500 years ago, and has an estimated dynamical lifetime (from integrations of the motion) of $\tau_d \sim$10$^4$ years \citep{Poz18}.  By comparison, the nucleus lifetime to physical decay may be much shorter.  
In particular, the rapidly changing spin seems likely to drive the nucleus of TGK to rotational breakup in short order (as suggested by \cite{Bod18}).  The  inverse period for breakup of a strengthless, spherical body is $\Omega_C = (G \rho_n/(3\pi))^{1/2}$, where $G = 6.67\times10^{-11}$ N kg$^{-2}$ m$^2$ is the gravitational constant.  Substituting $\rho_n$ = 500 kg m$^{-3}$ gives $\Omega_C = 6\times10^{-5}$ s$^{-1}$ (a rotation period of 4.7 hours).  The time to spin the nucleus from $\Omega$ = 0 to $\Omega_C$ at the rate $\dot{\Omega} = 7.6\times10^{-13}$ s$^{-2}$ (as measured from the 2017 perihelion data, c.f., Figure \ref{freq}) is $\tau_s = \Omega_C/\dot{\Omega}$.  Substitution gives $\tau_s \sim 4\times10^8$ s (about 13 years, or 2.5 orbits).   A similarly short timescale is indicated by the relation $\tau_s \sim 100 r_n^2$, with $r_n$ expressed in kilometers, as inferred from  a study of eight short-period comets having perihelia near 1 au \citep{Jew21}.  For TGK with $r_n$ = 0.5 km, this relation gives $\tau_s$ = 25 years (about 5 orbits).  If, as seems likely, the magnitude and direction of the torque evolve from orbit to orbit,  the approach to instability will be more like a random walk than a steady progression and the true spin-up times will be longer than given by the simple model.  Given what we know, however, it is clear that $\tau_s < \tau_d$; the lifetime of TGK to rotational instability is far shorter than the dynamical lifetime.

Given that $\tau_s < \tau_d$,  why does TGK still exist? Two possibilities are evident. First, TGK could be in an atypically active state. For example, perhaps sublimation is normally more thoroughly stifled by a refractory mantle than at present, reducing the mass loss rates and the associated torques, and so lengthening the timescale for spin-up, $\tau_s$.  We know that order of magnitude variations exist in the perihelion water production rates from orbit to orbit \citep{Com20}, making variable activity a plausible explanation for the survival of the nucleus. Second, TGK could have entered its present orbit as a larger nucleus, possibly of kilometer scale, and was then whittled down by sublimation or rotational instability to its current, diminutive form. The spin-up timescale varies as $r_n^4$ if the mass loss rate remains constant, or as $r_n^2$ if the rate scales in proportion to the nucleus surface area \citep{Jew21}.  In either case, a larger nucleus might survive against complete rotational breakup for a time $> \tau_d$, leaving the current nucleus as a remnant.  Unfortunately, we possess no evidence with which to firmly distinguish between these possibilities, nor to exclude their simultaneous action.

It is clear that the short timescales for variability make TGK an ideal test bed for the study of nucleus rotational dynamics.   Excited (NPA) rotation of the  nucleus seems quite likely as a result of its small size, but the degree of excitation to be expected is unclear and evidence from other small bodies shows that the observational signal of precession can be subtle.  Comet 1P/Halley famously shows evidence for NPA \citep{Sam91} but the better studied nucleus of 67P/Churyumov-Gerasimenko, for example, has a spin pole that is stable to $\lesssim$1\degr~even in the presence of substantial short-term variations in the rotational period caused by outgassing torques \citep{Gut16}.  More elaborate models of the rotation of TGK, including the possibility of non-principal axis rotation, will require much better data than we currently possess.
In this regard, we note that the observational geometry at the next perihelion (predicted for 2028 February 16) will be less favorable than in 2017 but we still expect that new observations will be of value in better understanding the spin state of the nucleus.   

\clearpage
\section{SUMMARY}
We present archival Hubble Space Telescope observations of Jupiter family comet 41P/Tuttle-Giacobini-Kresak from 2017, showing that its nucleus rotation period continued to change after perihelion, likely leading to a reversal of the spin. The observed, rapid changes are natural consequences of torques from outgassed volatiles acting on the very small nucleus.  Specifically, our results include

\begin{itemize}
\item The nucleus has radius $r_n$ = 0.5$\pm$0.1 km, a two-peaked rotation period  $P$ = 0.599 day (in 2017 December), and a lightcurve range $\Delta V =$ 0.4 magnitudes, indicating nucleus axes in the ratio $a/b \gtrsim$ 1.4:1. The dimensionless moment arm of the outgassing torque is $k_T$ = 0.013, about twice the median value measured in other Jupiter family comets.

\item The rotation of the nucleus likely slowed to zero in 2017 June and then reversed under the action of the torque.  

\item The active fraction decreased by an order of magnitude from $f_A \sim$ 2.4 in 2001 to 0.14 in 2017, indicating secular evolution of the outgassing surface on the orbital timescale.

\item The lifetime of the nucleus to rotational instability (a few decades) is short compared to the dynamical lifetime ($\sim10^3$ years) in its current orbit.  The continued existence of 41P therefore suggests that either the current level of outgassing activity is substantially larger than on average, and/or that the nucleus is a remnant of a once much larger body.

\end{itemize}

\acknowledgments
I thank Jane Luu, Yoonyoung Kim and the anonymous reviewer for comments on the paper.





\begin{deluxetable}{llccrrrrrrcrrrr}
\tablecaption{Hubble Observations of 41P
\label{geometry}}
\tablewidth{0pt}
\tablehead{\colhead{$t_{DOY}$\tablenotemark{a}} & \colhead{Date\tablenotemark{b}}  &  \colhead{$r_H$\tablenotemark{c}}   & \colhead{$\Delta$\tablenotemark{d}} & \colhead{$\alpha$\tablenotemark{e}}  & \colhead{$\theta_{- \odot}$\tablenotemark{f}} & \colhead{$\theta_{-V}$\tablenotemark{g}} & \colhead{$\delta_{\oplus}$\tablenotemark{h}} & \colhead{$\nu$\tablenotemark{i}}     }

\startdata

345 & Dec 11 & 2.784&  3.270&    16.3&   74.6& 261.9&    2.5&  124.7 \\
346 & Dec 12 & 2.792&  3.288&    16.1&   74.5& 261.8&    2.4&  124.9 \\
347 & Dec 13 & 2.799&  3.306&    15.9&   74.5& 261.7&    2.4&  125.0 \\
348 & Dec 14 & 2.806&  3.324&    15.7&   74.4& 261.6&    2.4&  125.2 \\

\enddata

\tablenotetext{a}{Day of Year, 1 = UT 2017 January 1}
\tablenotetext{b}{UT Date in 2017}
\tablenotetext{c}{Heliocentric distance, in au }
\tablenotetext{d}{Geocentric distance, in au }
\tablenotetext{e}{Phase angle, in degrees }
\tablenotetext{f}{Position angle of projected anti-solar direction, in degrees }
\tablenotetext{g}{Position angle of negative heliocentric velocity vector, in degrees}
\tablenotetext{h}{Angle of Earth from orbital plane, in degrees}
\tablenotetext{i}{True anomaly, in degrees}

\end{deluxetable}

\clearpage

\begin{deluxetable}{llllll}
\tablecaption{Photometry
\label{photometry}}
\tablewidth{0pt}
\tablehead{\colhead{N\tablenotemark{a}} & \colhead{UT Date\tablenotemark{b}} & \colhead{$V_5$\tablenotemark{c}} & \colhead{$H_{5}$\tablenotemark{d}} & \colhead{$V_{10}$\tablenotemark{e}} & \colhead{$H_{10}$\tablenotemark{f}}}
\startdata

1	&	11.193	&	24.18	&	18.72	&	23.82	&	18.36	\\
2	&	11.207	&	24.13	&	18.67	&	23.89	&	18.43	\\
3	&	11.533	&	24.11	&	18.65	&	23.93	&	18.47	\\
4	&	11.855	&	24.02	&	18.56	&	23.73	&	18.27	\\
5	&	11.870	&	23.98	&	18.52	&	23.80	&	18.34	\\
6	&	11.884	&	24.09	&	18.63	&	23.72	&	18.26	\\
7	&	12.120	&	24.17	&	18.71	&	23.81	&	18.35	\\
8	&	12.135	&	24.31	&	18.85	&	23.99	&	18.53	\\
9	&	12.149	&	24.14	&	18.68	&	23.92	&	18.46	\\
10	&	12.387	&	24.19	&	18.73	&	23.87	&	18.41	\\
11	&	12.402	&	24.02	&	18.56	&	23.25	&	17.79	\\
12	&	12.417	&	24.08	&	18.62	&	23.82	&	18.36	\\
13	&	12.731	&	24.08	&	18.62	&	23.82	&	18.36	\\
14	&	12.745	&	24.11	&	18.65	&	23.28	&	17.82	\\
15	&	13.113	&	24.15	&	18.69	&	24.02	&	18.56	\\
16	&	13.128	&	24.09	&	18.63	&	23.83	&	18.37	\\
17	&	13.449	&	24.40	&	18.94	&	24.05	&	18.59	\\
18	&	13.464	&	24.39	&	18.93	&	24.10	&	18.64	\\
19	&	13.516	&	24.46	&	19.00	&	24.00	&	18.54	\\
20	&	13.709	&	24.20	&	18.74	&	23.94	&	18.48	\\
21	&	13.739	&	24.25	&	18.79	&	23.93	&	18.47	\\
22	&	14.041	&	24.35	&	18.89	&	24.08	&	18.62	\\
23	&	14.055	&	24.26	&	18.80	&	24.01	&	18.55	\\
24	&	14.070	&	24.36	&	18.90	&	23.20	&	17.74	\\	

\enddata


\tablenotetext{a}{Running image number}
\tablenotetext{b}{UT Date in 2017 December}
\tablenotetext{c}{Apparent red magnitude within 5 pixel radius aperture}
\tablenotetext{d}{Absolute magnitude within 5 pixel radius aperture}
\tablenotetext{e}{Apparent red magnitude within 10 pixel radius aperture}
\tablenotetext{f}{Absolute magnitude within 10 pixel radius aperture}
\end{deluxetable}

\clearpage

\clearpage

\clearpage

\clearpage

\clearpage

\end{document}